\documentclass[twocolumn,pre,floats,aps,amsmath,amssymb]{revtex4}
\usepackage{graphicx,amsmath,subfigure}
\usepackage{bm,mhchem}
\usepackage{natbib}
\usepackage{array}
\newcolumntype{L}[1]{>{\raggedright\let\newline\\\arraybackslash\hspace{0pt}}m{#1}}
\newcolumntype{C}[1]{>{\centering\let\newline\\\arraybackslash\hspace{0pt}}m{#1}}
\newcolumntype{R}[1]{>{\raggedleft\let\newline\\\arraybackslash\hspace{0pt}}m{#1}}
\begin{document}

\title{Thin film detection of High Energy Materials:Optical Pumping Approach}
\author{Sachin Barthwal$^a$\footnote{corresponding author email:sachin.pranav@gmail.com}, Ashok Vudayagiri$^{a,b}$}
\affiliation{$^a$Advanced Centre for Research in High Energy Materials, University of Hyderabad \\$^b$School of Physics, University of Hyderabad, Hyderabad 500046, India}
\date{\today}

\begin{abstract}
We present our work on High Energy Material detection based on thin film of Lithium using the phenomenon of Optical Pumping. The Lithium atoms present in the thin film are optically pumped to one of the ground hyperfine energy levels so that they can no more absorb light from the resonant light source. Now in presence of a RF signal, which quantifies the ambient magnetic field, this polarised atomic system is again randomised thus making it reabsorb the resonant light. This gives a quantified measurement of the magnetic field surrounding the thin film detector. This is then mapped to the presence of magnetic HEMs and hence the HEMs are detected. Our approach in this regard starts with verifying the stability of Lithium atoms in various solvents so as to get a suitable liquid medium to form a thin film. In this regard, various UV-visible characterisation spectra are presented to finally approach a stable system for the detection. We have worked on around 10 polar and non polar solvents to see the stability criteria. UV-Visible probe has been found to be quite effective in addressing issue of hosting Lithium atoms by various solvents. Further results on the atomic system size as dispersed in the various solvents are presented. Atomic system size gives us a good quantitative estimate of the density of Lithium dispersed system in the thin film.  Finally a separate atomic system under Ultra-High-Vacuum is presented to achieve optical pumping wherein the polarisation and depolarisation of atomic system is seen using a 670 nm diode laser system which is electronically locked using a double window Hollow cathode lamp. These both systems will finally be combined together to give the final form to the detection system. As this whole system measures relative change in magnetic field due to the presence of HEMs, it gives a more reliable and accurate detection method compared to the absolute measurement detectors. \end{abstract} 

\maketitle
\section{Introduction} In the last decade or so, detection of High Energy Materials (HEMs) for security screening, detection of non exploded ordnance, for demining and pollution check have become an active and important area of research1-3. One main category of the detection of HEMs include the bulk detection systems wherein X Ray or neutron imaging are employed to locate large quantities of explosives4. Another method involves vapour and particle detection where trace quantities of explosives are detected via molecular signatures4. One of the main tools used for such detection is a LASER. Laser offers two main advantages in these detections due to their long range propagation of intense energy and high wavelength specificity with regards to atomic/molecular spectroscopy. 

Present work explores a somewhat new territory of thin film detection using a long known tool of optical pumping5. Thin film detection of HEMs have been recently explored elsewhere5 ,however in context of physical/absorption property of the thin film. In the present work magnetic property of the HEMs is used6 to detect them using the metal thin film magnetometer. Lithium atoms present in the thin film are optically pumped so that they can no more absorb light from the resonant light source. Now in presence of a RF signal, which quantifies the ambient magnetic field, this polarised atomic system is again randomised thus making it reabsorb the resonant light. This gives a quantified measurement of the magnetic field surrounding the thin film detector. This is then mapped to the presence of magnetic HEMs and hence the HEMs are detected. One measures the Larmor precession frequency $\omega$ of atomic spins in a magnetic field B, given by 

\begin{equation}
\omega= \gamma|B|
\end{equation}

where the gyro magnetic ratio $\gamma$ serves as the conversion factor between the frequency and the field strength.  As equation (1) suggests, the Larmor precession frequency measurement gives a measure of the B field around. So a measurement of Larmor precession frequency gives a measure of the ambient B field which is accomplished by using a RF field. This when mapped to the HEM magnetic field gives the final shape to the HEM detector.

 \begin{figure}[ht]
\includegraphics[width=3 in]{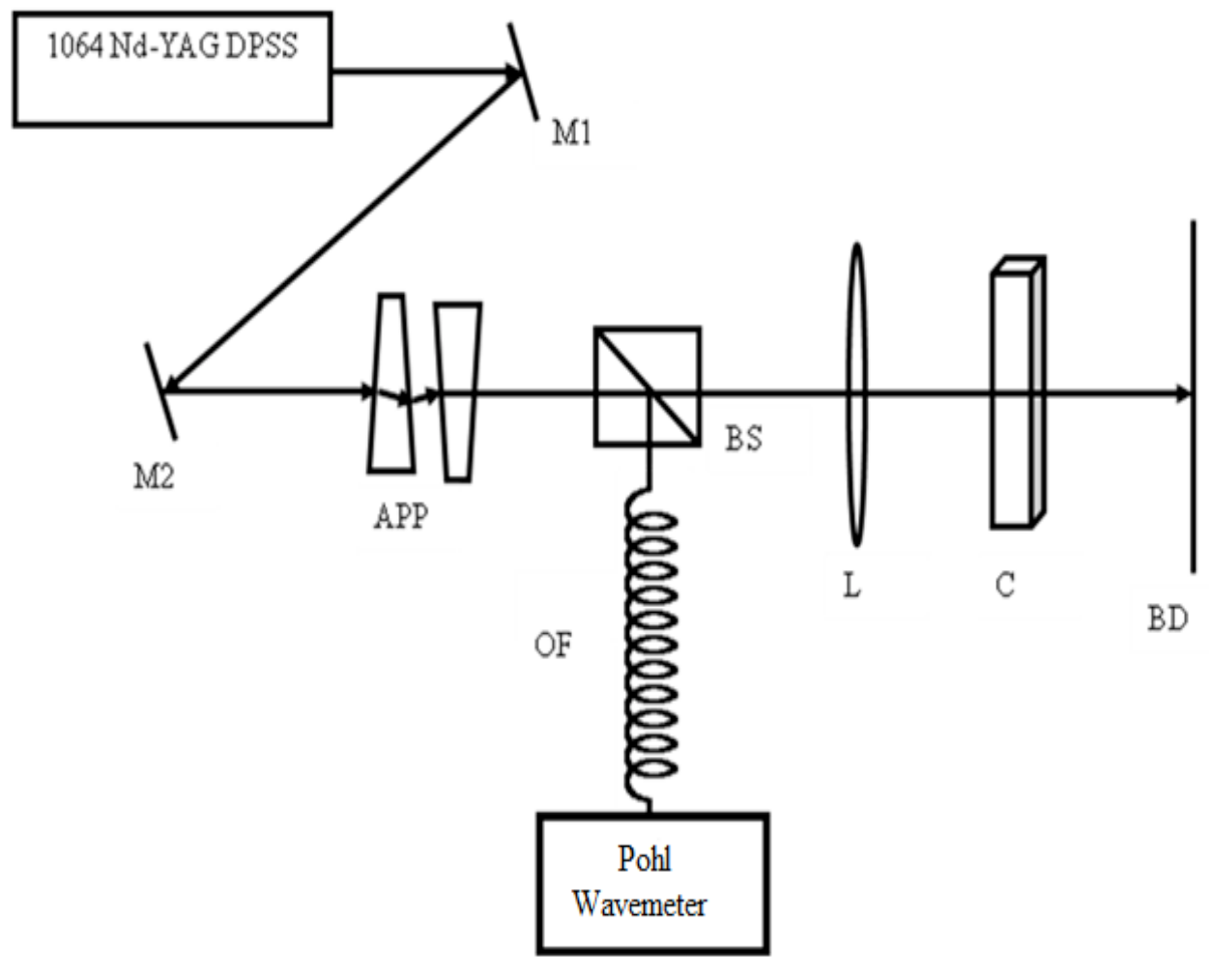}
\caption{Experimental set up showing excitation of Li in various solvents using DPSS laser. M1/M2: Steering mirrors, APP: Anamorphic prism pair, BS: Beam Splitter, OF: Optical Fibre, L: Lens, C: Cuvette, BD: Beam Dump. }
\label{fig: setup}
\end{figure}  
 
\section{Experimental Procedure}        

\subsection {Li stability with various solvents; UV-Vis spectroscopy of Li-solutions} 
First step in making the Lithium thin film is to check for the stability of Li in various solvents. We have done the same using UV/Visible Spectroscopy of Lithium in various polar/ non-polar solvents. This is done in presence/ absence of Laser beam irradiating the atomic system. Laser system is a 1064 nm DPSS Nd-YAG laser. The absorption spectra of irradiated Li solutions taken in the quartz cuvette of optical length10 mm were measured with wavelength from 200 to 800 nm using the Perkin Elmer UV-visible spectrophotometer. All the reagents and solvents of high pure grade were purchased from Sigma, Aldrich, Merck, Alfa-Aesar and used without further purifications. 1mM Li solutions using solvents have been prepared. The saturated solution of PMMA was prepared by dissolving the polymer in chloroform for 48 hrs at room temperature and then Li was added.

 \begin{figure}[ht]
\includegraphics[width=3 in]{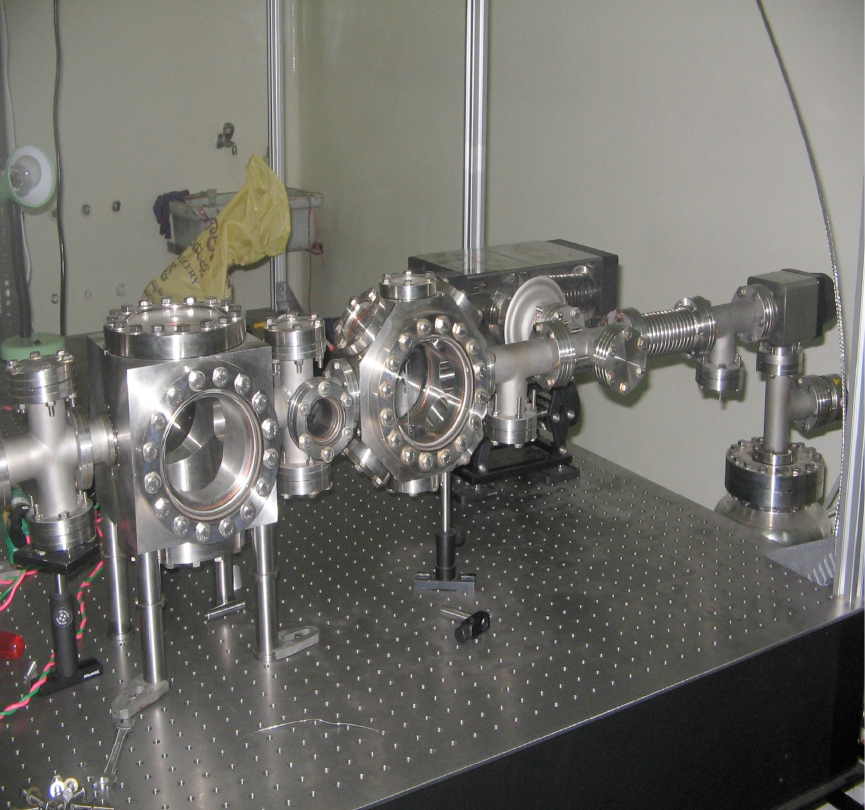}
\caption{Full UHV vacuum chamber showing the built atomic beam system  }
\label{fig: setup}
\end{figure}

\subsection{SEM Characterisation for the Li particles distribution in solvents}
 Lithium system with various solvents is then characterised with regards to the particle size dispersed in the solvents to estimate particle density in the resulting thin film. This is done using the SEM characterisation. This is important in making sure that Lithium particle are indeed uniformly distributed in the solvent and give a good system for performing optical pumping experiments.

 \begin{figure}[ht]
\includegraphics[width=2.5 in]{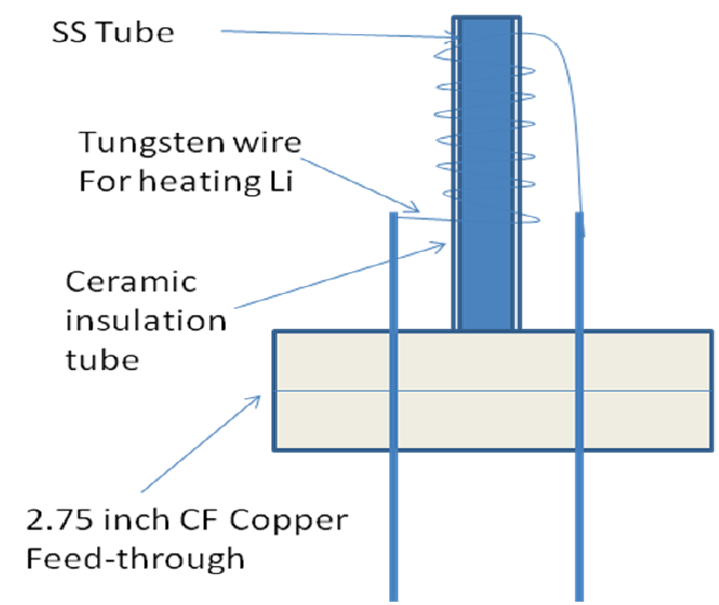}
\caption{Lithium source used for the experiment}
\label{fig: setup}
\end{figure} 

\subsection{Atomic beam system for achieving optical pumping in Li}
Another part of the experiment we plan to show optical pumping phenomenon in Lithium atomic beam system. For the same we successfully made a Li atomic beam setup using a combination of Rotary-Turbo-Ion pump to go down to ultra-high-vacuum (UHV) of ~10-9 torr of pressure.  Figure 3 shows full UHV chambered Lithium atomic beam.  The atomic source (Figure 4) basically comprises of a thin walled ss tube welded to a two pin feed through with solid Li inside which is resistively heated to around 4000C to get enough vapours.  This is achieved by passing around 4-5 A of current across the coil. To avoid divergence of the beam we have put an aperture, a circular Cu plate with a 1 mm hole punched at the centre of it. This gives a narrow Li atomic beam with transverse divergence of roughly 1 cm at around 10 cm away from the source. This was checked by tuning a diode laser at 670 nm to the resonance for the Li D lines and seeing the transverse extension of the fluorescence spot through a window. Typical fluorescence signal captured by a CCD camera (30 FPS Thor labs USB camera) are shown in the Figure 5. Laser used for the purpose is an External Cavity Diode Laser (ECDL) at 670 nm tunable from 665-675 from Toptica photonics with power output of 35 mW. 

\begin{figure}[ht]
\includegraphics[width=2.5 in]{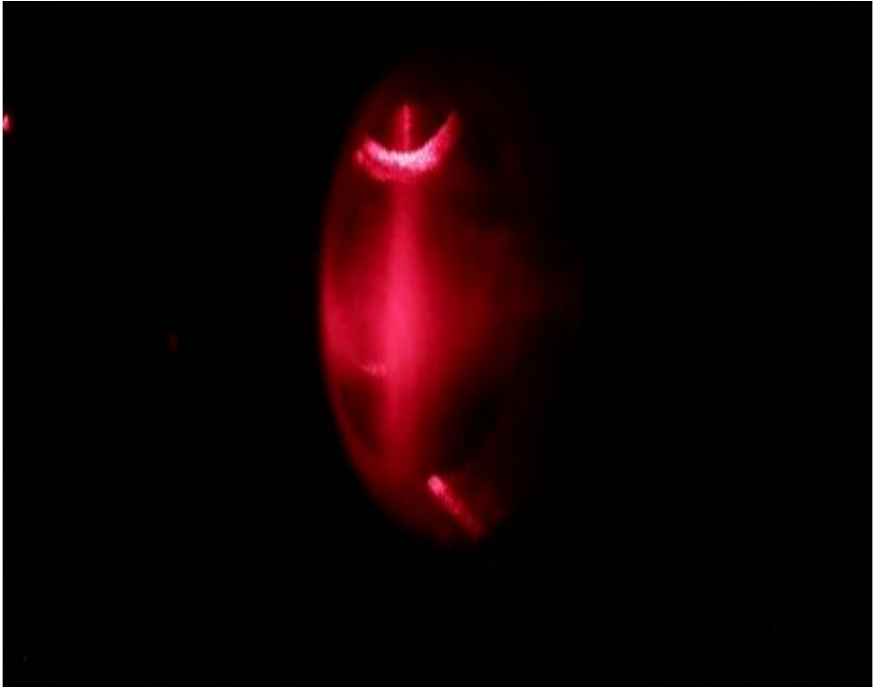}
\caption{Fluorescence signal as captured by a CCD camera from the Lithium atomic beam D line excited by ECDL at 670 nm }
\label{fig: setup}
\end{figure} 

\section{Results and Discussion}
\subsection{UV-Vis spectroscopy of Li-solutions}
 
The $\lambda_max$ of compounds is changed with the solvent and the concentration of solution. Four types of shifts in the $\lambda_{max}$ values are possible: (i) shift of $\lambda_{max}$ to longer wavelength or less energy is due to the substituent or solvent effect is referred as bathochromic or red shift (ii) shift of $\lambda_{max}$ to shorter wavelength or higher energy is due to the substituent or solvent effect is referred as hypsochromic or blue shift (iii) an increase in the absorption intensity called as hyperchromic shift and (iv) decrease in the absorption intensity called as hypochromic shift. The electronic transitions are usually accompanied by a simultaneous change between the vibrational levels. We found this shift happening in all the solvents. A typical spectra is shown in Figure 6 for acetonitrile and Lithium solution. 

\begin{figure}[ht]
\includegraphics[width=2.5 in]{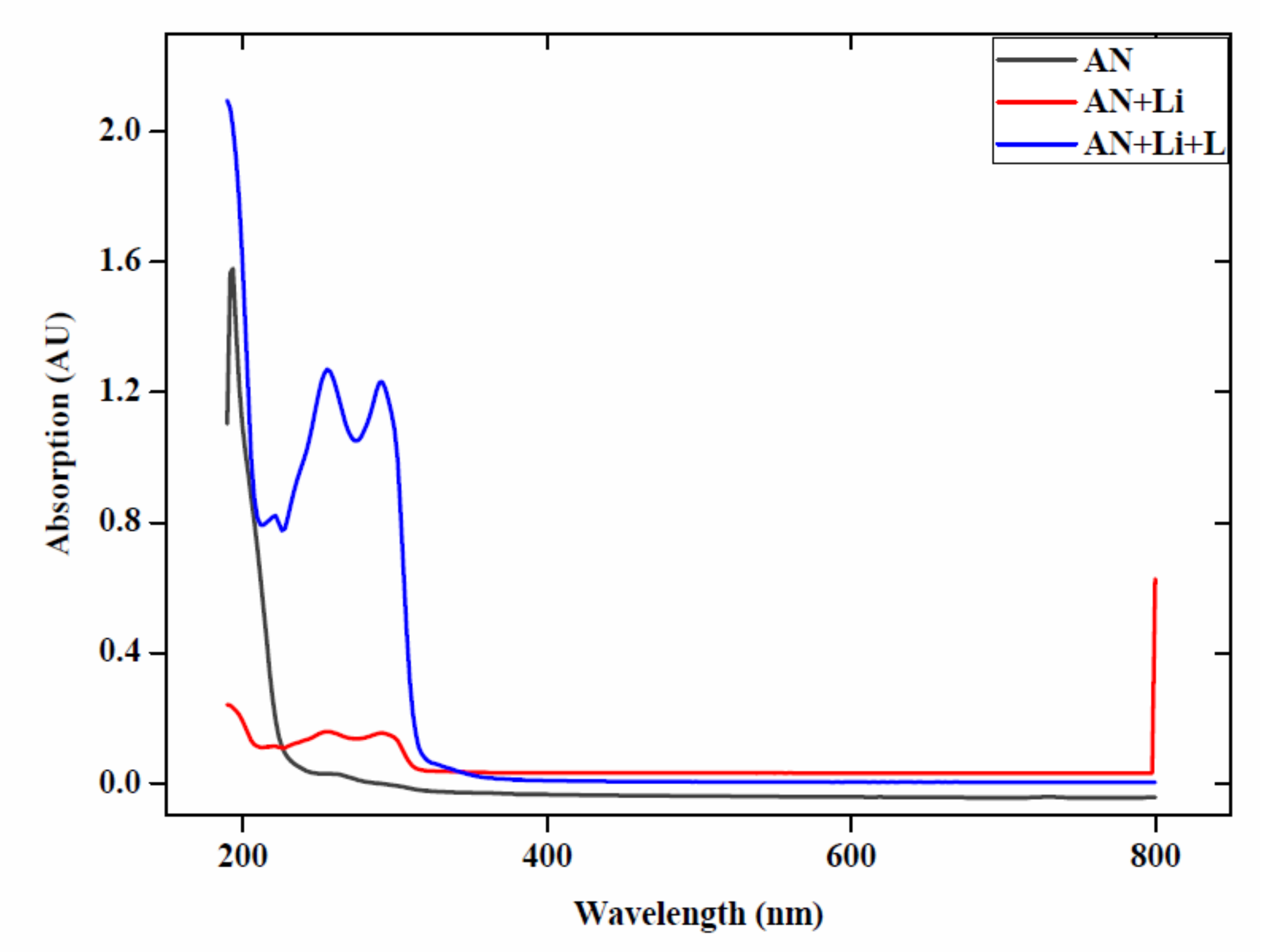}
\caption{UV-Visible spectra of Li-Acetonitrile solution with and without the laser irradiation (AN: Acetonitrile, L: Laser)}
\label{fig: setup}
\end{figure}  

Such spectra were taken for nine polar and non-polar solvents and the shift in cut off wavelength measured (Table 1). Except for the case of Acetonitrile (AN) we found the cut off wavelength varying post laser irradiation (Table 1). This gave us confidence that AN can be the perfect solvent with which Lithium can be added to form a stable film which will not get affected by laser perturbation and optical pumping can be performed with such combination. To give brief chemical reactivity of AN with Li, AN reacts with Li at room temperature to liberate hydrogen gas in Figure 6, the sharp band at 194 nm moved to the higher wavelength of 190 nm (i.e., red shift) and two new broad bands have appeared at 256 nm and 292 nm. It is due to the formation of mono-lithium derivatives of acetonitrile and released hydrogen gas on irradiation. The UV-visible spectrum of the Li solution shows cyanide absorption higher than the parent acetonitrile indicating a weakening of the cyanide linkage.

\begin{table*}
\caption{Various Solvents and their cut-off wavelengths found experimentally}
\begin{ruledtabular} 
\begin{flushleft}{
\begin{tabular}{llcccc}


Sl. & Solvent & {$\lambda_{max}$ (nm)  } & {$\lambda_{max}$ (nm) } &{ $\lambda_{max}$ (nm) } &{ $\lambda_{max}$ (nm) } \\ 
 & & literature & present study & Li solution pre irradiation & Li solution post irradiation\\ \hline
1 & Acetonitrile & 190 & 192& 256,292 & 256,292\\
2 & Hexane & 210 & 206,262 & 208,262 & 212,260 \\
3 & Methanol & 210 & 198 & 218 & 216,252 \\
4 & 1,4-Dioxane & 215 & 218,278 & 238,290 & 238,278\\
5 & Chloroform & 245 & 222 & 244 & 244,274\\
6 & Carbon Tetrachloride & 265 & 218 & 226,262, 294 & 264\\
7 & Toluene & 285 & 218 & 288 & 284\\
8 & Chlorobenzene & 285 & 232 & 288 & 288\\
9 & DMF & 270,300 & 224 & 274, 324, 350, 424 & 274,352, 388\\ 
\end{tabular}}
\end{flushleft}
\end{ruledtabular}
\end{table*}

The nitriles are known to undergo n $\to$ $\sigma*$ transition and n $\to$ $\pi*$ transition of characteristic cut off wavelengths of ~194 nm and ~300 nm. The two distinctive bands at 256 nm and 292 nm after irradiation are due to the formation of \ce{Li-CH_2CN}, an organometallic compound. The \ce{Li-CH_2CN} bond is of intermediate character between ionic and covalent bonds. The various bands of solutions both irradiated and non-irradiated apparently of the characteristic bands of lithium containing compounds: \ce{Li-CH_2CN} $\leftrightarrow$ \ce{Li-CH=C=NH} $\leftrightarrow$ \ce{Li-C$\equiv$CNH_2} $\leftrightarrow$ \ce{CH$\equiv$CN-Li-H} $\leftrightarrow$ \ce{CH_2-C=N-Li}. 

\begin{figure}[ht]
\includegraphics[width=2.5 in]{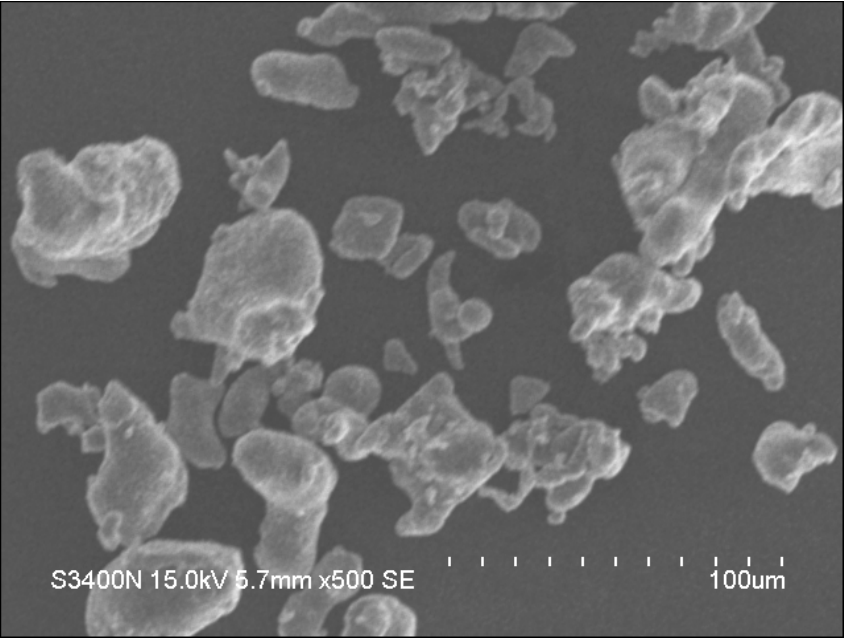}
\caption{SEM image of Li-AN solution showing uniformly distributed Li particles}
\label{fig: setup}
\end{figure}  

\subsection{SEM Characterisation for the Li particles distribution in solvents}
 
 \section{Conclusions and future aim}
Further SEM characterisation was conducted to ascertain the density of the Li particles in the solution with various solvents. Analysing the case of AN we found that Li particles get uniformly distributed in the solvent and give good system to work towards optical pumping of Lithium. 

\subsection{Atomic beam system for achieving optical pumping in Li}
 
Finally to achieve optical pumping in vapour system we worked on confining the UHV Li atomic beam in two directions using laser cooling. Post image analysis results are as shown in the Figure  below.

 \begin{figure}[ht]
\includegraphics[width=2.5 in]{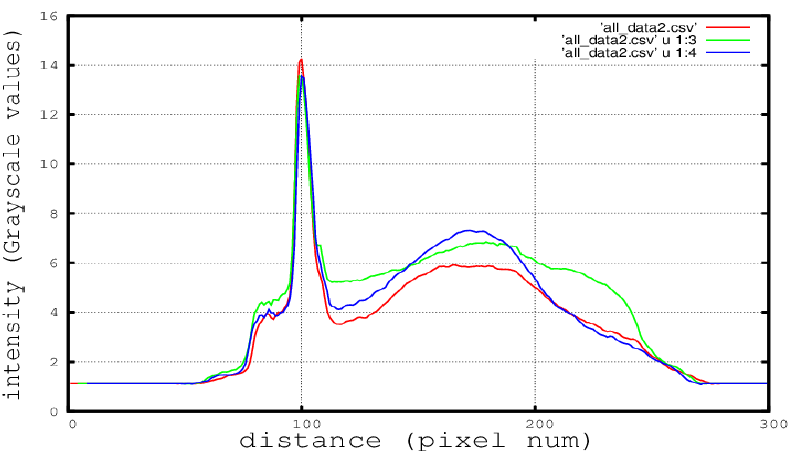}
\caption{Beam collimation happening with laser going near resonance }
\label{fig: setup}
\end{figure}  
 \section{Conclusions and future aim} 

In conclusion we can say that we have worked towards making a HEM detection system using phenomenon of optical pumping. We first investigated stability of Li in various solvents and found AN as the most suitable one to make a stable thin film. This system we analysed for Lithium distribution using SEM and found the distribution as uniform. Before going for optical pumping on this system we made a UHV Li beam at $10^{-9}$ Torr pressure wherein beam as confined using laser cooling to achieve optical pumping. As future work we plan to achieve optical pumping in the thin film system and finally use it for detecting presence of HEMs. \\

\section{Acknowledgement}
 
National fusion program, Board of research on fusion science and technology and DRDO is greatly acknowledged for the financial assistance for the present work.

\end{document}